\documentclass[11pt,twoside]{article}
\usepackage{cspm2015}

\aspSuppressVolSlug
\resetcounters

\newcommand{\rsun}{R$_\odot$}
\newcommand{\kmps}{kms$^{-1}$}

\begin{document}

\title{Interplanetary Consequences of Coronal Mass Ejection Events 
occurred during 18--25 June 2015}

%\author{P. K. Manoharan                - below list of authors modified on 30 JAN 2016
%\affil{Radio Astronomy Centre, National Centre for Radio Astrophysics,
%Tata Institute of Fundamental Research, 
%Udhagamandalam (Ooty) 643001, India; \email{mano@ncra.tifr.res.in}}}

\author{P.K.~Manoharan,$^1$ D.~Maia,$^2$ A.~Johri,$^1$ and M.S.~Induja$^1$
\affil{$^1$Radio Astronomy Centre, National Centre for Radio Astrophysics,
           Tata Institute of Fundamental Research, 
           Udhagamandalam (Ooty) 643001, India; \email{mano@ncra.tifr.res.in}}
\affil{$^2$CICGE, Faculdade de Ciencias da Universidade do Porto, Porto, Portugal} }

% This section is for ADS Processing.  There must be one line per author.
\paperauthor{P.K. Manoharan}{mano@ncra.tifr.res.in}{}{Tata Institute of Fundamental Research}{Radio Astronomy Centre, National Centre for Radio Astrophysics}{Udhagamandalam (Ooty)}{Tamilnadu}{643001}{India}
\paperauthor{D. Maia}{dmaia@fc.up.pt}{}{CICGE}{Faculdade de Ciencias da Universidade do Porto}{Porto}{}{}{Portugal}
\paperauthor{A. Johri}{johri@ncra.tifr.res.in}{}{Tata Institute of Fundamental Research}{Radio Astronomy Centre, National Centre for Radio Astrophysics}{Udhagamandalam (Ooty)}{Tamilnadu}{643001}{India}
\paperauthor{M.S. Induja}{msindu89@gmail.com}{}{Tata Institute of Fundamental Research}{Radio Astronomy Centre, National Centre for Radio Astrophysics}{Udhagamandalam (Ooty)}{Tamilnadu}{643001}{India}

\begin{abstract}
In this paper, we review the preliminary results on the propagation effects 
and interplanetary consequences of fast and wide coronal mass ejection (CME) 
events, occurred during 
18--25 June 2015, in the Sun-Earth distance range. The interplanetary scintillation 
(IPS) images reveal that the large-scale structures of CME-driven disturbances 
filled nearly the entire inner heliosphere with a range of speeds, $\sim$300--1000 
{\kmps}. The comparison of speed data sets, from IPS technique results in the inner 
heliosphere and {\it in-situ} measurements at 1 AU, indicates that the drag 
force imposed by the low-speed wind dominated heliosphere on the propagation 
of CMEs may not be effective. The arrival of shocks at 1 AU 
suggests that a shock can be driven in the interplanetary medium  by the 
central part of the moving CME and also by a different part away from its
centre.  The increased flux of proton at energies $>$10 MeV is consistent 
with the acceleration of particles by the shock ahead of the CME.

\end{abstract}

%\section{Introduction}

%In the current solar cycle, some of the CME events occurred during June 
%2015 were fast as well as wide, drove shocks, and caused an intense 
%geomagnetic storm. We track the plasma characteristics of these CME 
%events as they propagate into the interplanetary medium.

\section{Introduction -- Coronal Mass Ejections during 18--25 June 2015}

Coronal mass ejections (CMEs) move outward from the Sun into the solar
wind. Understanding the effects of CMEs in the Sun-Earth distance and 
identification of their interacting shocks at the Earth's magnetosphere 
are scientifically important as well as essential for space weather 
perspectives. Numerous studies, including models, have been made to 
investigate the propagation of CMEs in the inner heliosphere 
(e.g., \cite{2004SpWea...2.9001D};
\cite{2001ApJ...559.1180M}; 
\cite{2010SoPh..265...49B}).
%Dryer et al. 2004; Manoharan et al 2001; Manoharan 2010).
In this study, we consider the radial evolution of large CMEs
occurred in the current solar cycle.

According to the sunspot observations, the current solar cycle \#24
manifests to be a less intense one in comparison with the previous 
recent cycles (in fact, this is the smallest sunspot cycle since 
cycle \#14, \url{http://solarscience.msfc.nasa.gov}). In the course 
of the current cycle, however, several small and large sunspot groups 
have produced moderately intense flare/CME events. During June 2015, 
two groups of fairly large flare-active sunspots passed across the 
visible disk of the Sun: one in the southern hemisphere (AR\#2367, 
with a magnetic classification of $\beta\gamma$) and the other in the 
northern hemisphere (AR\#2371, $\beta\gamma\delta$ classification). 
In this study, we consider the period, 18--25 June 2015, during which 
a number of flares/CMEs and their related eruptions from the above
active regions propagated through the heliosphere. In particular, 
five large CMEs were observed from the above active regions. In 
Table~1, we list the timings and locations of these five CME 
events, along with the available speeds of type-II radio bursts.
Four of them were fast halo
CMEs (i.e., V$_{\rm CME}$ $>$1000 {\kmps}) and were associated with 
M-class flares at AR\#2371. The second event in the list originated
from AR\#2367 at S21W27 and propelled a partial halo CME of width 
$>$150 degrees, at a nominal speed of $\sim$300 {\kmps}.

\begin{table}[!ht]
%\caption{List of flares, CMEs, and type-II radio bursts} - modified on 30 JAN 2016
\caption{List of flares, CMEs, and type-II radio bursts$^\dagger$}

\smallskip
\begin{center}
\begin{tabular}{clllllllc} % define the column alignment
                           % l: left, c: center, r: right
\hline                     % horizontal line
No.& Date  & Flare    & Flare  & CME   & CME   & CME       & Type-II    & Active \\
   &(2015) & Class    & Loc.   & Time  & Type  & Speed     & Speed      & Region \\
   &       &          &        & (UT)  &       &({\kmps})  & ({\kmps})  & \#     \\
\hline
1 & 18 June & M3.0/1N & N15E50 & 17:24 & Halo  &$\sim$1100 &            & 2371 \\
2 & 19 June &         & S21W27 & 06:42 & PHalo & $\sim$300 &            & 2367 \\
3 & 21 June & M2.6    & N12E13 & 02:48 & Halo  &$\sim$1200 & $>$700     & 2371 \\
4 & 22 June & M6.5    & N12W08 & 18:36 & Halo  &  1250     & $\sim$1400 & 2371 \\
5 & 25 June & M7.9/3B & N10W42 & 08:36 & Halo  &  1500     & $\sim$1700 & 2371 \\
\hline
\end{tabular}
\end{center}
%below footnote included on 30 JAN 2016
\footnotesize{$^\dagger$The solar radio spectral data sets are from 
US Air Force Radio Solar Telescope Network (RSTN) and are available at
\url{ftp://ftp.ngdc.noaa.gov/STP/SOLAR\_DATA}}. 

\end{table}

\begin{figure}[b]
\centering
\includegraphics[width=0.65\textwidth]{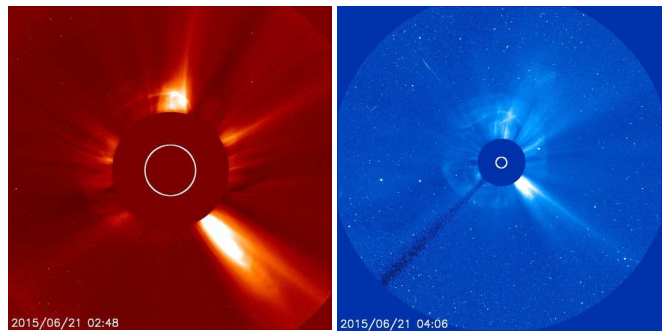}
\caption{LASCO white-light images of the CME event on 21 June 2015. 
The {\it left} and {\it right} images are, respectively, from 
C2 and C3 coronagraphs, observed at 02:48 and 04:06 UT. The
white circle indicates the limb of the Sun. The colored figure 
can be found in the electronic version.}
\end{figure}

A major geomagnetic storm (Dst = -204 nT for a prolonged period on 
23 June 2015; \url{http://wdc.kugi.kyoto-u.ac.jp}) 
was caused by the fast CME event on 21 June 2015 (Table~1, event \#3), 
which was associated with an M2.6 flare from AR\#2371 at N12E13. The 
white-light images of the above CME event, obtained from the LASCO 
C2/C3 coronagraphs 
\cite{1995SoPh..162..357B},
%(Brueckner et al. 1995), 
are displayed in Figure 1. 
The fast expanding full halo structure, along with the filament moving 
in the north-east direction, can evidently be observed in these images. 
Since the originating location of the CME event 
was to the east of the central meridian of the Sun, the direction of 
propagation of the CME (also its expansion) was oriented to the 
east with respect to the Sun-Earth line. The LASCO images typically
cover up to a heliocentric distance of about 30 solar radii ({\rsun}) 
(1 AU $\approx$ 215 {\rsun}). The interplanetary scintillation (IPS) 
observations have been employed in tracking the consequences of 
propagation of CMEs farther into the inner heliosphere at distances 
$\geq$50 {\rsun} (e.g., 
\cite{2000ApJ...530.1061M};
\cite{2001ApJ...559.1180M}).
%Manoharan et al. 2000 and 2001). 

%\articlefigure[width=.65\textwidth]{figure_1.eps}{figure_1}
%{LASCO white-light images of the CME event on 21 June 2015. 
%The {\it left} and {\it right} images are, respectively, from 
%C2 and C3 coronagraphs, observed at 02:48 and 04:06 UT. The
%white circle indicates the limb of the Sun. The colored figure 
%can be found in the electronic version.}

%\articlefigure[width=1.0\textwidth]{figure_2.eps}{figure_2}
%{Ooty IPS images for the period 19--26 June 2015. These are similar 
%to the LASCO images and show the sky-plane projection of the inner heliosphere.
%The concentric circles are of radii, 50, 100, 150, and 200 {\rsun}. The 
%red color code indicates the background solar wind. In these images, 
%observing time increases from right (west of Sun) to left (east of Sun).
%The enhanced scintillation regions are the CME-driven disturbances in 
%the IPS field of view. 
%The white patches indicate the data gaps. %included on 30 JAN 2016
%The colored figure can be found in the electronic version.}

\begin{figure}[t]
\centering
\includegraphics[width=1.0\textwidth]{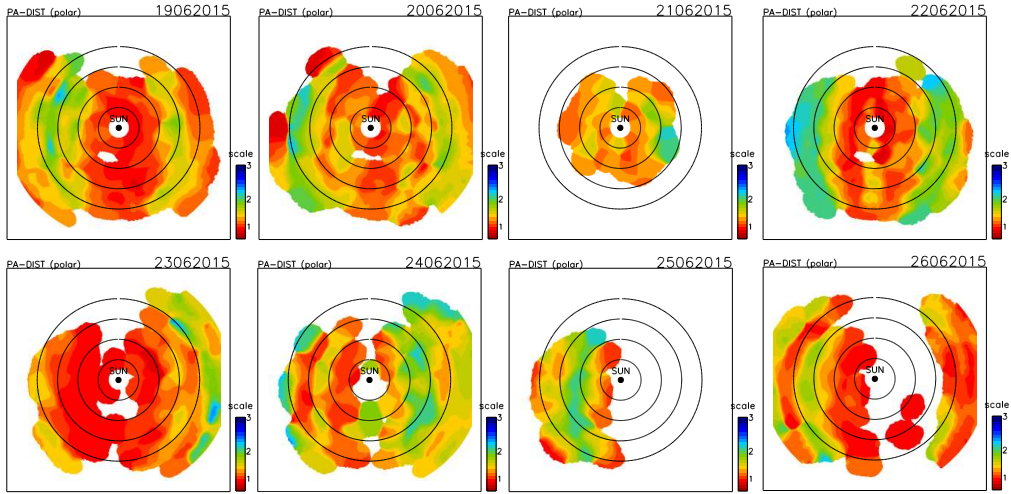}
\caption{Ooty IPS images for the period 19--26 June 2015. These are similar 
to the LASCO images and show the sky-plane projection of the inner heliosphere.
The concentric circles are of radii, 50, 100, 150, and 200 {\rsun}. The 
red color code indicates the background solar wind. In these images, 
observing time increases from right (west of Sun) to left (east of Sun).
The enhanced scintillation regions are the CME-driven disturbances in 
the IPS field of view. 
The white patches indicate the data gaps. %included on 30 JAN 2016
The colored figure can be found in the electronic version.}
\end{figure}

\markboth{P.K. Manoharan et al.}{Interplanetary Consequences of CMEs}

\subsection{Interplanetary Scintillation Images}

The IPS observation from a radio source can provide 
estimates of solar wind speed and normalized scintillation index 
({\it g}-index, which is directly related to the level of density 
turbulence in the solar wind, i.e., $\Delta$N$_{\rm e}$) at the 
closest solar approach of the line of sight to the radio source 
(e.g.,
\cite{1993SoPh..148..153M};
\cite{2000ApJ...530.1061M}).
%(e.g., Manoharan 1993; Manoharan et al. 2000).
The IPS measurements at the Ooty Radio Telescope (ORT) on a grid of a large 
number of radio sources allow to image the disturbances associated 
with the CMEs at different distances from the Sun before their arrival 
at the near-Earth space (for details on Ooty IPS studies refer to 
\cite{2000ApJ...530.1061M};
\cite{2010SoPh..265..137M};
\cite{2010SoPh..265...49B}).
%Manoharan et al. 2000, 2001; Manoharan 2010; Bisi et al. 2010).  
The images of normalized scintillation index ({\it g}-maps) obtained from 
the Ooty measurements between 19 and 26 June 2015 are displayed in 
Figure 2. These images are the sky-plane projection of the inner 
heliosphere, as viewed from the Earth and the vertical (top to 
bottom) and horizontal (left to right) directions, respectively, 
represent north-south and east-west directions of the heliosphere. 
In these images, the observing time increases from right to left
(i.e., from west of the Sun to east). Thus, each image shows the 
three-dimensional distribution of $\Delta$N$_{\rm e}$ on a given 
day around the Sun, over a heliospheric diameter of $\sim$2 AU. 
For example, a {\it g}-value close to unity, $g$~$\approx$~1, 
corresponds 
to the condition of the ambient solar wind and values of g~$>$~1 and 
g~$<$~1 represent, respectively, enhancement and depletion in levels 
of density turbulence (i.e., $\Delta$N$_{\rm e}$) in the solar wind.
The large-scale structures of enhanced scintillation seen in these
images show the presence of CME-associated disturbances in the 
interplanetary medium. The outward displacement from the centre 
of the heliosphere and increase in size of these structures with 
time suggest the typical speed of propagation and interplanetary 
consequences of the CMEs.

In this period of study, 18--25 June 2015, the Sun was not dominated 
by mid-latitude coronal hole(s) and the contribution from the interaction
of fast-slow solar wind streams was expected to be insignificant  
(in general, the effects of such interactions are observed at distances 
larger than 1 AU (e.g., 
\cite{1996ARA&A..34...35G})). 
%Gosling 1996)).  
Therefore, the IPS images 
shown in Figure 2 reveal the consequences of CMEs related disturbances in 
the interplanetary medium. On all these days, the inner heliosphere was
filled with the CME-associated disturbances of a high level of density 
turbulence.

\section{Speed Distribution of CME-driven Disturbances}

Some of the disturbances seen in the IPS images were produced by the 
fast halo CMEs and they moved with speeds higher than that of the speed 
of the ambient solar wind. Solar wind speed estimates obtained from
Ooty IPS observations between 19 and 27 June 2015 are plotted in 
Figure 3. These plots include speed data from sources for which the 
observed {\it g}-values are well above the ambient level (i.e., 
$g$~$\geq$~2) and confirm speeds associated with the CME disturbances.
It is to be mentioned that since IPS technique probes different parts
of the heliosphere, the above plots would include speed measurements
from various parts of the CME-driven disturbances.
Moreover, since we are interested 
in understanding the effects of propagation of the CMEs with distance 
from the Sun, we compare the speeds of solar wind at
two heliocentric distance ranges, respectively, $\leq$125 {\rsun} and 
$>$125 {\rsun}. It is evident in these plots that during the passage of 
CMEs, the interplanetary medium is filled with a range of speeds, 
$\sim$300--1000 {\kmps}. During this period, as indicated by the 
IPS observations as well as {\it in-situ} measurements at 1 AU
(refer to Figures 4 and 5), the speed of the ambient (i.e., 
background) solar wind varies in the range of $\sim$275--375 {\kmps}. The 
drag applied by the background solar wind (i.e., the interaction 
of the CME with the ambient flow) is proportional to 
$|{\rm V_{CME} - V_{AMBIENT}}|^2$. Therefore, in the current 
situation (i.e., low-speed dominated background solar wind) the CMEs would likely be slowed down.
However, as seen in the above plots,
the evolution of high-speed wind (i.e., $>$500 {\kmps}) 
between these two regions of the inner heliosphere (i.e., with 
respect to the midway between the Sun and Earth) is not significant. 
A similar speed distribution is observed at both the distance 
ranges.

\articlefigure[width=0.40\textwidth,angle=-90]{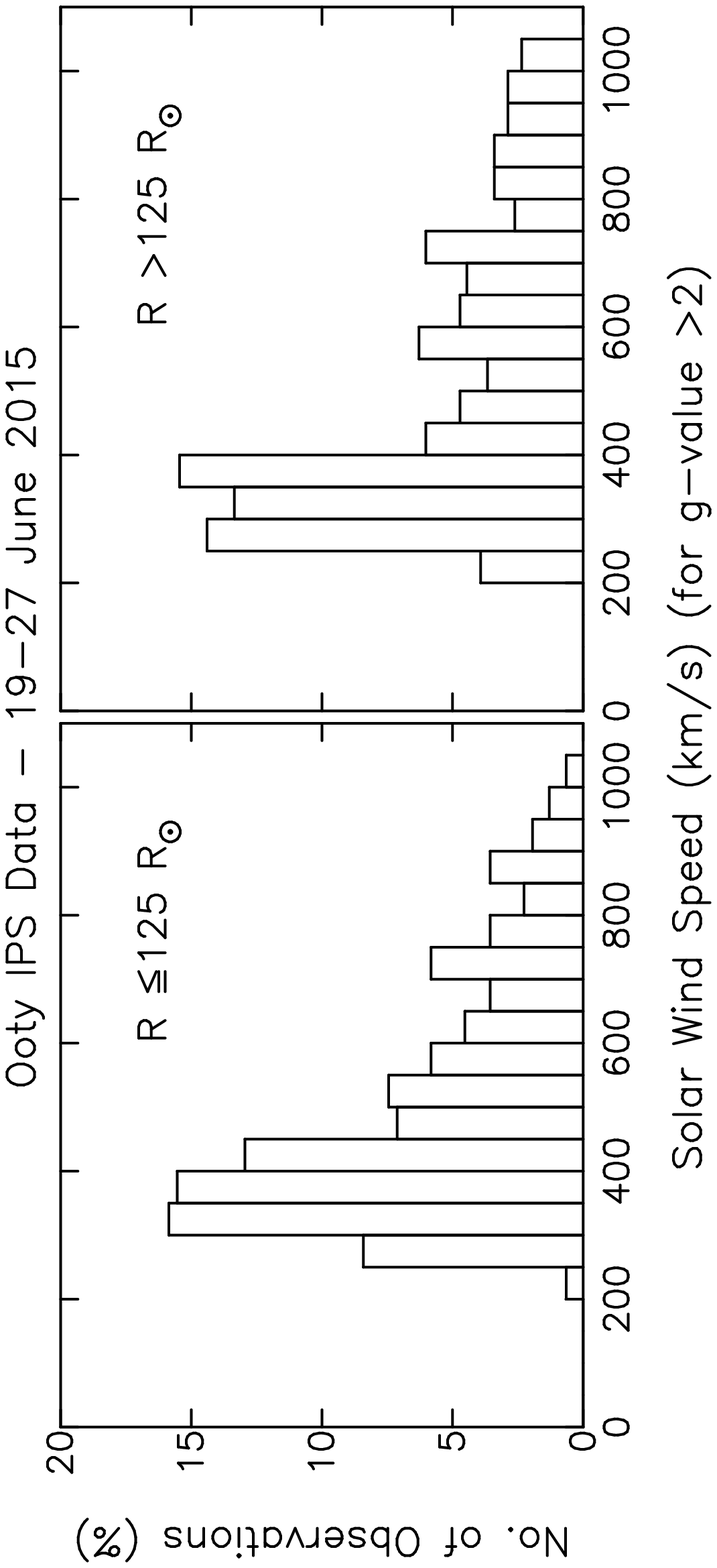}{figure_3}
{Histograms of solar wind speed estimates obtained from the Ooty
IPS measurements at distances, respectively, $\leq$125 {\rsun} 
({\it left} plot) and $>$125 {\rsun} ({\it right} plot). The $\it y-$axis 
marks the percentage of number of observations in each speed bin normalized by 
the total number of observations considered. These plots include speed 
data from sources for which the measured g-values are $\geq$2 and 
this criterion would allow only the enhanced level of scintillation caused by 
the propagating CME disturbances to be taken into account.}

\articlefigure[width=0.85\textwidth]{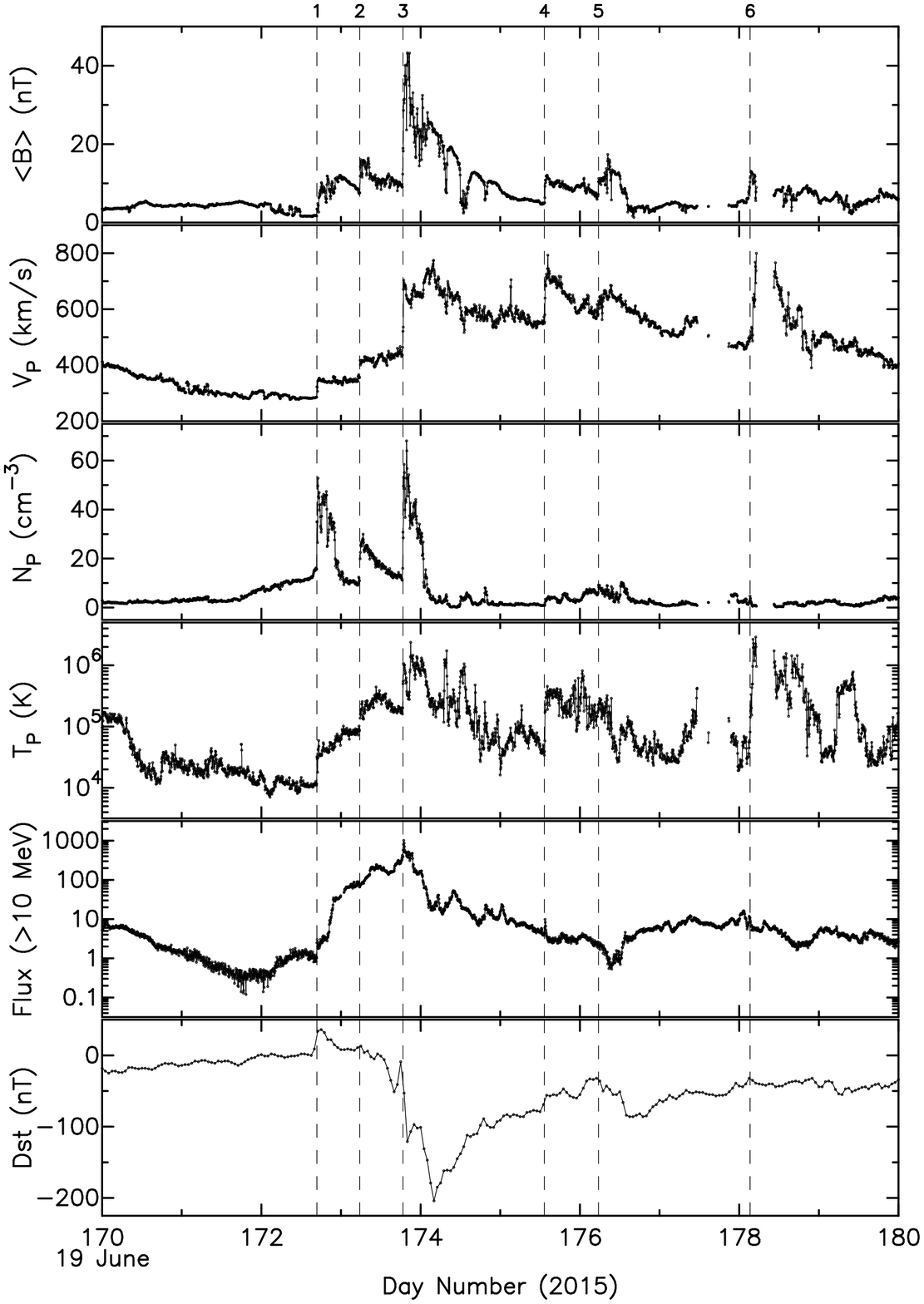}{figure_4}
{Interplanetary magnetic field and plasma parameters for days
170 to 180 (19--28 June 2015). The geomagnetic index, Dst value,
is plotted in the bottom panel of the figure. The Dst index is 
hourly averaged data and other data sets are from 5-min resolution 
measurements.  The vertical dotted lines indicate the arrival
times of shocks and shock numbers are shown at the top.}

\subsection{Interplanetary Magnetic Field and Plasma Parameters}

It is to be noted that the CMEs seen in the IPS field of view were 
not earthward directed (Figure 2 and refer to Table~1) and thus the 
centre part of the 
CMEs would not have been expected to hit the Earth and cause effects 
at the near-Earth space. However, in this period of the study, six
interplanetary shocks were recorded in the {\it in-situ} data sets 
observed at the near-Earth orbit and two geomagnetic storms (i.e.,
an intense storm, Dst = -204 nT and the other of moderate strength
storm, Dst = -86 nT) were observed by the geomagnetic observatories.
Figure 4 shows the interplanetary plasma and magnetic field parameters 
of the ambient solar wind and CME-driven shock disturbances observed 
at 1 AU. These data sets have been obtained from the {\it OMNIWeb Plus 
Interface} (\url{http://omniweb.gsfc.nasa.gov}). 
Figure 4 shows from top to bottom (i) average magnetic field, $<$B$>$,
(ii) solar wind proton speed, V$_{\rm P}$, (iii) density, N$_{\rm P}$, 
(iv) proton temperature, T$_{\rm P}$, and (v) proton flux at energy $>$10 
MeV. These plots have been made using 5-min averaged data sets. Figure 4 
also includes the plot of hourly Dst value, which denotes the intensity of 
the geomagnetic storm. This figure illustrates the conditions of the
solar wind prior to the shocks (i.e., unshocked ambient solar wind),
shocked plasma and the conditions of driver gas associated with the 
CMEs. The arrival of six shocks are indicated by dashed vertical lines. 
The CME behind the shock \#3 revealed a flux-rope like structure in
the plasma and magnetic field data. The proton flux plotted in Figure 4 
shows increase in flux between $\sim$21 and 28 June 2015. It is likely 
that the above CME-driven shocks in the Sun-Earth space have supported 
% to accelerate particles to high energies. - below line modified on 30 JAN 2016
the acceleration of particles to high energies.

Since the IPS observations allow to probe the properties of the solar wind 
in three-dimensional space, the preliminary results of Ooty IPS tomographic 
reconstruction suggest that (i) CME events \#3 and \#4 were nearly Earth 
directed and the earlier one caused an intense shock as well as a major 
storm of intensity Dst = -204 nT at the Earth (refer to shock \#3 in Figure 4); 
(ii) for other events listed in Table~1, only the tail part of the CME crossed 
the Earth (i.e., just a glancing blow at the Earth's magnetosphere) and caused 
shocks of considerable strength, but not the intense storms. The reconstruction
also shows that some of the CMEs move intact and retain their loop-like shape 
even at large solar distances, suggesting that the magnetic energy associated 
with a CME is important in determining its radial evolution. Given the 
limitation of space in this book, a detailed paper on the study of IPS 
tomographic reconstruction of these CME events will be published elsewhere.

\articlefigure[width=0.40\textwidth,angle=-90]{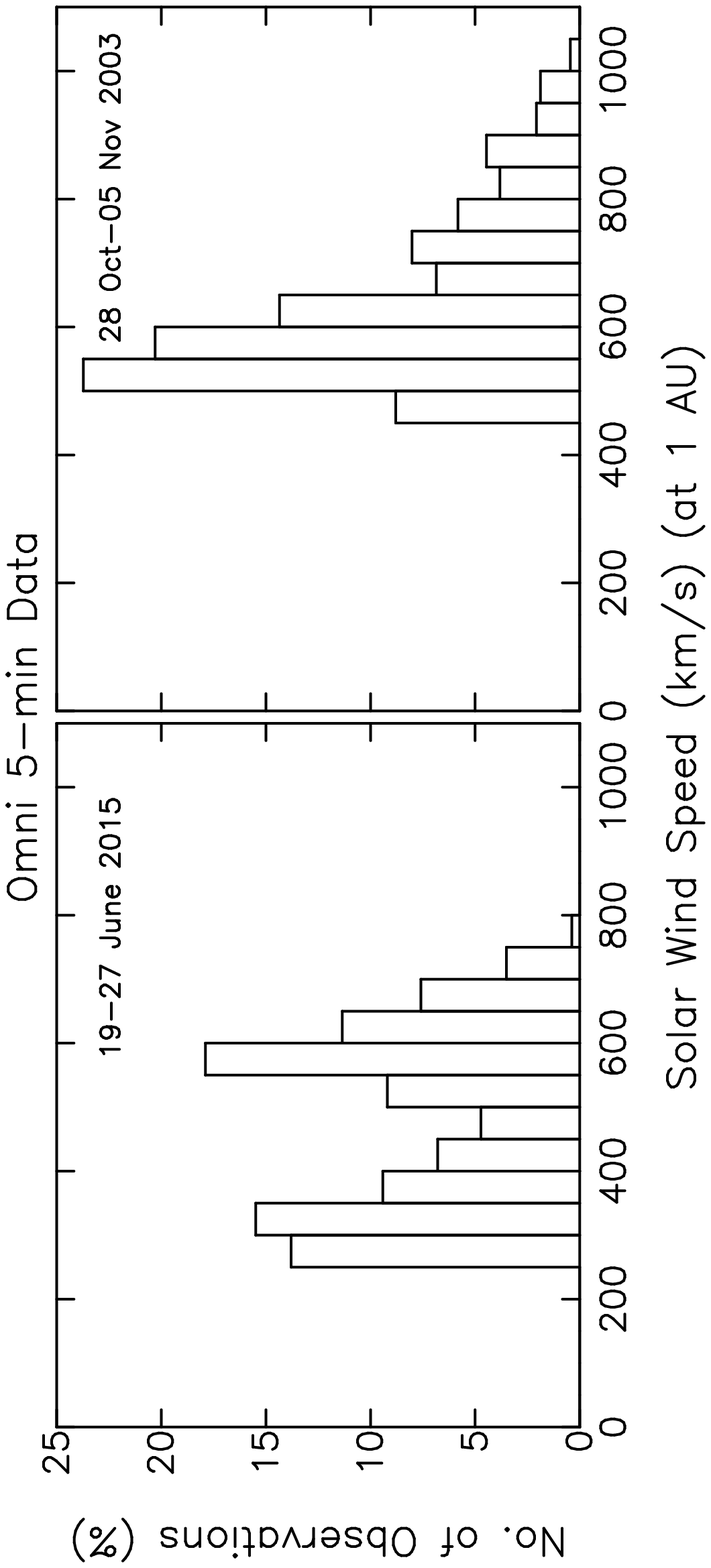}{figure_5}
{Histograms of {\it in-situ} solar wind speed obtained from
the OMNI database (\url{http://omniweb.gsfc.nasa.gov/}),
for the time periods, respectively, 19--27 June 2015 
({\it left} plot) and 28 October--05 November 2003 ({\it right}
plot). These plots include ambient solar wind conditions
as well as disturbances caused by the CMEs at the near-Earth
space.}

In Figure 4, it is to be noted that during 22--26 June 2015 (i.e., day 
numbers 173 to 178), the solar wind speed at 1 AU remains at a high
level, which is well above the ambient speed. When we examine the speed 
distribution, obtained from the {\it in-situ} data at 1 AU, it shows 
that a significant fraction of data points is in the high speed range 
$>$500 {\kmps}. Figure 5 ({\it left} panel) shows the histogram of solar 
wind speed data obtained from the {\it in-situ} measurements for the period
19--27 June 2015. It is to be mentioned that this plot includes speeds 
from the ambient solar wind and CME-associated disturbances. Since not
all CME-driven disturbances are Earth directed, this plot includes
propagation characteristics of the nearly central part of the CME crossing
the Earth as well as the effects of the east or the west wings of the CMEs. As stated
earlier, the plots made from the IPS observations would include speed 
data from various parts of the propagating CME-driven disturbances.
The speed of the ambient (i.e., background) solar wind seems to be
low at below 350 {\kmps}, whereas the speed distribution associated 
with the CME-driven disturbances lies much above the background flow. 
At $\sim$1 AU, the drag caused by the ambient solar wind has not been
efficient enough to slow down the CME disturbances. It is consistent
with the range of speeds obtained from the IPS technique, which gives 
an overall view of the propagating structures and a significant fraction 
of them travel at high speeds.

The above inferred solar wind conditions, observed during 19--27 June
2015, have been compared with the largest storm period of the nearly
similar phase of the previous cycle, i.e., 28 October -- 05 November
2003 (Halloween solar events period). The {\it right} panel of Figure 5
shows the histogram of solar wind speed data from the {\it in-situ} 
measurements. In this plot, the striking feature is that the background solar wind
speed is above $\sim$350 {\kmps} and is different from the low-speed
dominated heliosphere of the current cycle. In spite of the low 
background speed, in the period 19--27 June 2015, the CME-driven 
disturbances continue to travel at speeds above 500 {\kmps}. Whereas
in the Halloween period, most of the CMEs originated with rather high 
speeds (i.e., $\sim$1000--2000 {\kmps}) and they significantly evolved 
in the Sun-Earth distance.

\section{Summary}

The interplanetary consequences of CME events occurred during 18--25 June
2015 have been analysed based on the white-light images in the near-Sun region, 
IPS images covering a heliocentric distance range of 50--200 {\rsun}, and 
{\it in-situ} measurements at 1 AU. During this period, the inner heliosphere
was filled with CME-driven disturbances of a high level of density turbulence
($\Delta {\rm N_e}$) and their speeds covered a wide range, $\sim$300--1000 
{\kmps}. Moreover, at this phase of the cycle, the heliosphere was dominated 
by low-speed ambient flows ($\sim$275--375 {\kmps}), which likely did not
slow down the CMEs. The solar wind speed distributions at two distances (i.e., 
$\leq$125 {\rsun} and $>$125 {\rsun}) as well as at 1 AU are consistent with 
the less radial evolution of speed. The loop-like structures seen in some 
of the CMEs maintain their shape even at large distances from the Sun and it
indicates the importance of magnetic and/or internal energy within a CME in 
controlling the radial evolution (e.g., 
\cite{2010SoPh..265..137M};
\cite{2011JASTP..73..671M}).
%Manoharan 2010; Manoharan and Rahman 2011). 
Since the effective
radial evolution can vary from one CME to the other, it is essential that not 
only the drag effected by the background flows, but also the internal energy of a CME
and its rate of dissipation in the interplanetary medium are required to be
considered in understanding the radial evolution of the CME.

\acknowledgements 
Members of the Radio Astronomy Centre (NCRA-TIFR) are acknowledged for 
making the ORT available for IPS observations. SOHO/LASCO is a project 
of international cooperation between ESA and NASA. The near-Earth solar 
wind data and geomagnetic indexes have been obtained from OMNIWeb service 
and OMNI data of NASA/GSFC's Space Physics Data Facility 
(\url{http://omniweb.gsfc.nasa.gov}). We acknowledge National Geophysical 
Data Center for the solar data used in this study. This work was partially 
supported by the CAWSES-India Program, sponsored by ISRO.

\bibliography{mano_astroph}

\end{document}